# Cyber Security in Containerization Platforms: A Comparative Study of Security Challenges, Measures and Best Practices

Sohome Adhikari[1] and Sabur Baidya[2]
[1]IT Shared Services, Yum! Brands, Louisville, KY, USA
[2]Department of Computer Science and Engineering, University of Louisville, KY, USA
Email: sohome.adhikari@yum.com, sabur.baidya@louisville.edu

*Abstract*— The paper reviews the comparative study of security measures, challenges, and best practices with a view to enhancing cyber safety in containerized platforms. This review is intended to give insight into the enhanced security posture of containerized environments, with a view to examining safety vulnerabilities in containerization platforms, exploring strategies for increasing containers isolation and assessing how encryption techniques play an important role in providing secure applications. The paper also provides practical guidance for organizations seeking to strengthen their cyber security defenses in the containerization area platforms.

## I. INTRODUCTION

In the rapidly evolving landscape of digital technology, containerization platforms have become increasingly prevalent for deploying and managing applications due to their efficiency and scalability. However, with the rise of containerization also comes the heightened importance of cyber security measures to safeguard sensitive data and protect against potential threats. This paper delves into enhancing cyber security in containerization platforms through a comparative study of security measures and best practices. By exploring the common security vulnerabilities in containerization platforms, examining strategies to enhance container isolation, and evaluating the role of encryption techniques in securing containerized applications, this research aims to provide insights into bolstering the security posture of containerized environments. Furthermore, by investigating recommended best practices for securing container images, exploring methods for implementing network security within container environments, and outlining critical considerations for access control and authentication in containerization platforms, this study seeks to offer practical guidance for organizations looking to fortify their cyber security defenses in the realm of containerization. Through a comprehensive analysis of security measures and best practices, this research paper aims to contribute to the ongoing dialogue on enhancing cyber security in containerization platforms and mitigating risks associated with containerized applications.

## II. IDENTIFYING CHALLENGES IN CONTAINERIZATION CYBERSECURITY

### A. What are the common vulnerabilities associated with containerization?

Containerization, while offering numerous benefits in software development and deployment, is not without its vulnerabilities that malicious actors can exploit. The fact that containerization does not necessarily eliminate vulnerabilities is one of the main points necessitating organizations to implement robust vulnerability management practices to ensure the security of their container ecosystems. These vulnerabilities can manifest in various forms, including attacks against container images, authentication mechanisms, applications, and network vulnerabilities [12][13]. A common challenge associated with container security is the tendency for developers to adopt a "set and forget" mentality during container development, potentially leaving room for vulnerabilities [8]. Security misconfigurations, such as default insecure configurations, further compound the risks in container environments and can serve as entry points for malicious activities [8]. To address these vulnerabilities effectively, organizations must prioritize secure container configurations, continuous vulnerability scanning, monitoring, and leveraging container security platforms with comprehensive threat detection capabilities [12]. Maintaining strong access controls, authentication mechanisms, and effective isolation techniques are essential strategies for mitigating unauthorized interactions and safeguarding containerized environments against potential threats [12]. In general, in order to ensure the confidentiality, integrity and availability of containerized applications and information at a time when security challenges are becoming more complex, it is important to remain vigilant by continuously scanning vulnerabilities, patch management as well as proactively detecting threats. [12].

### B. How do attackers exploit these vulnerabilities in container environments?

In the realm of container environments, attackers have a myriad of vulnerabilities at their disposal to exploit and breach security measures. For instance, attackers often target network vulnerabilities or misconfigurations within the container environment as a gateway to access sensitive data and systems [14]. Moreover, by exploiting vulnerabilities in container runtimes, attackers can seize control of containers, potentially compromising the entire container ecosystem [11]. This can extend to gaining root privileges on the host system by exploiting vulnerabilities in the container runtime, providing attackers with a significant foothold in the infrastructure [14]. In addition, attackers may manipulate unpatched vulnerabilities in commonly used applications within containers to infiltrate the environment and extract valuable information [14]. By breaching the registry hosting container images, malicious actors can inject harmful code into the containers, leading to the dissemination of malware across the containerized applications [11]. Furthermore, attackers can leverage shared Linux kernels in container environments to compromise one container and propagate their control to other interconnected containers, amplifying the scope of the breach significantly [15]. As organizations increasingly adopt emerging technologies and orchestration platforms, attackers are likely to focus on exploring potential vulnerabilities in these novel environments, necessitating a proactive approach towards security measures [15]. Ultimately, addressing these vulnerabilities and enhancing security measures through continuous monitoring, timely patching, and rigorous audits can help organizations mitigate the risks posed by malicious actors in container environments [14].

*C. What are the challenges in detecting and mitigating security gaps in containerization?*

Containerization presents unique challenges in terms of security, primarily due to the dynamic and decentralized nature of containerized runtime environments. Misconfigurations in container orchestration platforms can lead to security gaps, posing significant risks to organizations relying on container technology for their applications. The challenge lies in the potential for misconfigurations to introduce vulnerabilities that attackers can exploit to gain unauthorized access to container environments, emphasizing the importance of effective orchestration and configuration management in container security [15]. In order to meet these challenges, organizations need to put in place rigorous security measures including vulnerability scanning, configuration management, access control, network segmentation and monitoring with a view to protecting themselves against possible threats. Moreover, detecting and mitigating security gaps in containerization necessitates adherence to best practices and leveraging specialized security tools to enhance the security posture of containerized environments. One of the critical hurdles in container security is the limited visibility containers provide to developers, IT engineers, and security analysts into workloads, making threat detection and response more challenging. As containerized workloads involve managing a larger attack surface, organizations must stay proactive and comprehensive in their security approach to mitigate security gaps and protect against potential breaches [12]. Additionally, ensuring the validation of container environment configurations is crucial for preventing oversights that could increase the risk of security breaches [11]. In essence, addressing the complexities of container environments and staying informed about emerging threats while embracing technological advancements and industry best practices are essential strategies for effectively detecting and mitigating security gaps in containerization [12].

### III. IDENTIFYING MEASURES IN CONTAINERIZATION CYBERSECURITY

*A. How can container isolation be enhanced to prevent security breaches?*

Enhancing container isolation is crucial in preventing security breaches in containerized environments. By utilizing container security solutions, organizations can bolster the security of their containers against advanced threats and vulnerabilities, ultimately minimizing the risk of breaches in the system [4]. Platforms like CloudDefense offer robust container security solutions that enhance container isolation and protect against potential security risks. Implementing best practices and specialized security tools, such as containerized next-generation firewalls and microsegmentation tools, can help organizations safeguard their container environment from unauthorized access and data breaches while ensuring compliance with industry regulations [5]. Running containers with the least privileges is essential. Containers should not be run as the root user. Use security profiles like Seccomp and AppArmor to restrict the system calls a container can make.

Container orchestration platforms like Kubernetes manage container lifecycles. Access to the orchestration API with solid authentication and authorization mechanisms is crucial. Implement policies that define what a container can do within a pod. Define network policies within orchestration platforms to control communication between pods. Properly configuring security contexts in Kubernetes and enforcing the principle of least privilege are essential steps in enhancing container isolation and preventing security breaches [5].

Additionally, establishing firewalls, developing VPC rules, and isolating container networks are effective measures to prevent the spread of threats within the network and enhance container security [4][6]. Furthermore, monitoring containers during runtime is critical to prevent attacks like privilege escalation and lateral movement, contributing to overall container security. Ultimately, the host OS in a containerized environment significantly provides maximum container isolation and prevents security breaches. Implementing robust container isolation measures is essential in defending the container deployment environment and ensuring the security of containerized applications and infrastructure.

Security measures during runtime are also essential, like deploying runtime security tools that monitor container activities for malicious behavior. Many cloud providers give users tools to monitor the runtime data with their version of orchestration platforms. They also let users define runtime policies that can detect and respond to suspicious activities within containers.

*B. What role do encryption techniques play in securing containerized applications?*

Encryption techniques are indispensable in fortifying containerized applications against security breaches and unauthorized access. By encrypting sensitive information such as passwords, API keys, and tokens, encryption techniques provide a crucial layer of security to safeguard valuable data within containers [6]. Secrets management tools further enhance this security by encrypting and delivering secrets to containers as required, ensuring that sensitive information remains protected from prying eyes [6]. Adopting container-first development has brought about new security challenges, making encryption techniques even more critical in securing containerized applications [4]. Securing containers, including their associated images and code, can be complex, underscoring the importance of encryption in protecting sensitive data from potential vulnerabilities and cyber threats [4]. In light of the various vulnerabilities that containers are susceptible to, encryption techniques are vital in fortifying the security posture of containerized applications and mitigating risks associated with unauthorized access.

IV. IDENTIFYING BEST PRACTICES IN CONTAINERIZATION CYBERSECURITY

*A. What are the recommended best practices for securing container images?*

For the security of container images, following a set of recommended best practices is imperative. These practices start with the understanding that container images serve as the foundation of containers, encapsulating the essential components like code, runtime, libraries, and configurations required for application execution [6]. The first step in fortifying container security is to conduct thorough vulnerability scans on all images before deploying them [7]. It is crucial to rely only on reputable and minimal container images from trusted sources to mitigate security risks [7]. Regularly updating images with the latest security patches and ensuring they are free from vulnerabilities are essential steps in enhancing security measures [6].

It is recommended that container images be stored in secure private or public registries and their integrity be verified by checking image signatures using tools like Notary [2][8]. Securing container registries is paramount to guarantee team members leverage images devoid of vulnerabilities [2]. Additionally, continuous scanning of container images for vulnerabilities, misconfigurations, and secret keys is advised to maintain a robust security posture [7]. It is essential to integrate scanning tools into the CI/CD pipeline to automatically assess images for security risks during the build and deployment processes [9]. Ultimately, securing container images involves the images themselves and the entire stack used for building, distributing, and executing containers, emphasizing a comprehensive approach to container security [10].

*B. How can organizations implement network security within container environments?*

Implementing network security within container environments is a multifaceted task that demands comprehensive strategies and solutions. To begin with, organizations must prioritize robust container security practices to fortify their network security within container environments [9]. It involves a holistic approach encompassing container-environment networks to create a robust security posture [9]. Additionally, organizations must choose a security solution that can effectively cover various Kubernetes environments, including cloud-managed and self-managed setups, serverless containers, and standalone containers on virtual machines [7]. This solution should be able to discover and scan containers, hosts, and clusters across different Kubernetes environments to ensure comprehensive security coverage [7].

Moreover, continuously utilizing security controls is imperative to safeguard containerized environments from evolving security risks [8]. Network security within container environments extends beyond just securing containers and applications. It reinforces the fundamental elements, such as the container runtime, kernel, and host operating system, to establish a resilient security framework [8]. More than merely securing the host is required to provide complete protection in container environments, emphasizing the need for additional security measures [2]. Proper isolation between containers plays a pivotal role in enhancing network security within container environments, preventing unauthorized access and potential threats from spreading within the network [2][6].

Furthermore, maintaining secure configurations to restrict container permissions is essential to limit potential security breaches in container environments, underscoring the importance of proactive security measures [2]. Limiting inter-container communication is another critical aspect that organizations must prioritize to bolster network security within container environments, mitigating the risk of lateral movement of threats within the network [6]. Also, different resources can be put into separate resource groups, providing an additional layer of security.

*C. What are the key considerations for access control and authentication in containerization platforms?*

Access control and authentication are fundamental aspects of ensuring the security and integrity of containerization platforms. By implementing robust access control measures, only authorized users can interact with containers and sensitive data, preventing unauthorized access and potential security breaches [6]. Access control is critical in container security, limiting access to only necessary individuals and reducing the risk of unauthorized entry [6][9]. Authentication mechanisms are essential in verifying the identity of users and ensuring that only authorized personnel can access containers within the platform [9]. Furthermore, securing the underlying infrastructure and orchestration layer in containerization platforms is vital for overall security. Cloud providers offer tools and services to bolster the security of these components, safeguarding against vulnerabilities that could lead to unauthorized access or denial of service attacks [7]. Users must also take proactive steps to secure their containers and the applications they host within the platform, including securing components of the host system such as the operating system, libraries, and container runtime to mitigate security risks [7][10]. Securing the host system is paramount in preventing unauthorized access to running containers within the platform, emphasizing the importance of avoiding running containers as root unless necessary and utilizing user namespaces to isolate container privileges [10][6]. A zero trust approach is gradually replacing the traditional perimeter based safety model. It foresees increased security, by continuously verifying the user's identity and device integrity regardless of their location or network[17].

Additionally, implementing security contexts and securing the container engine are crucial steps in controlling access and capabilities within the containerization platform, ensuring containers are separated from each other and the host system [6][7]. Engineers should also enforce strict access controls in container registries to prevent unauthorized entry and monitor these registries for any suspicious access patterns that could indicate malicious activity within the platform [11]. Promptly addressing container security risks is essential for maintaining cloud-native security and minimizing security and compliance vulnerabilities that could impede cloud adoption [10].

### D. How can organizations enhance monitoring and incident response for container security?

To enhance monitoring and incident response for container security, organizations must adopt a multifaceted approach incorporating various strategies and tools. Embracing technological advancements is crucial in this endeavor, as it allows organizations to keep up to date with emerging threats and continually strengthen their security measures[12]. Leveraging industry best practices is another essential aspect that organizations should prioritize to enhance monitoring and incident response for container security [12]. *Continuous Monitoring* is a valuable strategy that organizations can adopt to continuously assess container vulnerabilities and anomalies, ensuring the integrity and resilience of containerized applications throughout their operational lifecycle [15]. By implementing Continuous Monitoring, organizations can promptly identify and respond to security threats for containerized applications, mitigating risks and reducing vulnerabilities [15][13].

Furthermore, visibility into the running information of both containers and hosts is critical for enhancing Monitoring for container security, as it enables security teams to effectively identify and mitigate vulnerabilities in containerized environments [8]. Additionally, leveraging the right tools, such as Check Point CloudGuard IaaS for virtual patching and integrating security tooling throughout the software development lifecycle (SDLC) and Continuous Integration/Continuous Deployment (CI/CD) pipelines are instrumental in improving incident response for container security [16]. By addressing challenges related to visibility into container workloads and adopting a holistic approach to Monitoring, organizations can strengthen their security posture and effectively respond to security incidents in containerized environments.

### E. What role does automation play in addressing cybersecurity gaps in containerization?

In cybersecurity within containerization, automation is critical for addressing and remedying existing security gaps. Automation is instrumental in enforcing continuous monitoring, swift response mechanisms, and proactive risk mitigation strategies in container security, ensuring that potential vulnerabilities are promptly identified and rectified [15]. By integrating automation and AI-driven solutions, evolving container escape vulnerabilities can be effectively tackled and mitigated [15]. The role of automation becomes even more pronounced in staying ahead of cybersecurity challenges specific to containerization, where consistent monitoring, logging, and regular audits are paramount for maintaining a secure and resilient environment [15][14]. Furthermore, automation aids in the timely application of patches to keep platforms and infrastructure up to date, a crucial aspect in fortifying containerized applications against cyber threats [14]. One of the key advantages of automation is streamlining security processes, ensuring that security measures are consistently applied across various elements within container environments, such as images, containers, hosts, runtimes, registries, and orchestration platforms [14]. Given the added complexity of multiple layers of abstraction in containerization, automation becomes indispensable in addressing cybersecurity gaps, offering a systematic approach to security enforcement and vulnerability scanning [14][12]. Moreover, specialized tools are essential for interpreting, monitoring, and safeguarding containerized environments, underscoring the necessity of automation in cybersecurity to manage and protect against potential risks [14] effectively.

### V. CONCLUSION

This paper delves into the critical aspects of enhancing cyber security within containerization platforms through a comprehensive comparative study of security measures and best practices. The discussion highlights the importance of addressing common security vulnerabilities in container environments to ensure a secure network environment. By leveraging tools like software composition analysis and binary analysis, organizations can proactively scan for vulnerabilities, detect malware, and enhance security measures within container platforms. Emphasizing the need for robust container isolation, the study underscores the significance of monitoring the software supply chain and implementing secure configurations to mitigate security risks effectively. Furthermore, encryption techniques, secure configurations in Kubernetes, and access control measures play pivotal roles in fortifying containerized applications against security breaches and unauthorized access. The discussion advocates for proactive security measures such as securing components of the host system, continuous utilization of security controls, and avoiding running containers as root to enhance container security. It also stresses the importance of a holistic approach encompassing securing container images, runtimes, networks, access control, monitoring, and logging to address vulnerabilities comprehensively. Recognizing the evolving nature of security risks, the discussion underscores the necessity of proactive security management to safeguard containerized environments effectively. Overall, the research paper contributes valuable insights into enhancing cyber security in containerization platforms and provides a roadmap for organizations to strengthen their security posture and minimize potential risks within containerized environments.